# Black phosphorus – a new saturable absorber material for ultrashort pulse generation

Jaroslaw Sotor[1,*], Grzegorz Sobon[1], Wojciech Macherzynski[2], Piotr Paletko[2], Krzysztof M. Abramski[1]

[1]*Laser & Fiber Electronics Group, Wroclaw University of Technology, Wybrzeze Wyspianskiego 27, 50-370 Wroclaw, Poland*
[2]*Faculty of Microsystem Electronics and Photonics, Wroclaw University of Technology, Janiszewskiego 11/17, Wroclaw 50-372, Poland*
[*]*jaroslaw.sotor@pwr.edu.pl*

**Abstract**

Low-dimensional materials, due to their versatile properties are very interesting for numerous electronics and optoelectronics applications. Recently rediscovered black phosphorus, with a graphite-like structure can be exfoliated up to the single atomic layer. In contrary to graphene it possesses a direct band gap controllable by the number of stacked atomic layers. For those reasons, it is now intensively investigated. Here we demonstrate, that black phosphorus can serve as a broadband saturable absorber and can be used for ultrashort optical pulse generation. The mechanically exfoliated ~300 nm thick layers of black phosphorus were transferred onto the fiber core and under pulsed excitation at 1560 nm wavelength its transmission increases by 4.4%. It was used to generate 272 fs-short pulses at 1550 nm and 739 fs at 1910 nm. The obtained results shows that black phosphorus can be effectively used for ultrashort pulse generation and proves its great potential to future applications.

Tremendous increase in the number of practical applications requiring ultrafast lasers[1-7] operating in the visible-to-mid-infrared spectral range stimulates the instantaneous progress in research on saturable absorbers (SA) for passive mode-locking of lasers[8-17]. Currently, most of the commercially available laser systems utilize semiconductor saturable absorber mirrors (SESAMs), thanks to their well-established technology and possibility to achieve desired optical parameters (modulation depth, saturation fluence and non-saturable losses)[8-10]. However, the SESAMs also have some limitations. Their operation bandwidth is relatively narrow because of the present band-gap of the used semiconductors. Moreover, the complicated and expensive molecular beam epitaxy (MBE) process essential in its fabrication is challenging in the mid-infrared spectral range (above 2 µm). All these factors caused development of a new class of broadband SAs based on carbon nanomaterials (carbon nanotubes[11,13,18-21], graphene[11-13,21-30]), topological insulators[14,15] and transition metal dichalcogenides[16,17,31]. The ultrashort pulses have been generated in fiber[11-15,17-19,21-24,31], solid state[20,21,25,26-28] and semiconductor lasers[29,30] in the 800 – 2400 nm spectral range using these nanomaterial-based SAs.

Among these new materials, graphene is the most versatile and is very attractive for electronics and optoelectronics[32-34]. Commonly considered to be a "wonderful material", besides the application in mode-locked lasers can be successfully used in many optoelectronic devices like: photodetectors[35], light modulators[36], polarizers[37], sensors[38], solar cells (transparent electrodes)[39], RF electronics[40], transistors[41] etc. Its highly application potential allows to think on integrated optoelectronic devices[42]. However, lack of the band gap in pristine graphene[43] is particularly unwanted in some electronic applications. Thus, it is often required to open the bandgap using complicated techniques[44-46]. These limitations lead to the intensive research on other two-dimensional materials which can complement graphene.

Recently the research community is focused on black phosphorous which is considered as a new promising material for optoelectronic applications[47]. Its structure is similar to graphite, where the following monolayers are bound with van der Walls forces[48,49]. Mechanical exfoliation of black phosphorus[50] allows to obtain a two-dimensional material called phosphorene[51], analogously to graphene. In terms of electronic applications, the most important feature is that black phosphorus is a direct bandgap semiconductor in the mono-, few and bulk form. Its band gap is scaled with the number of layer form ~1.5 eV for phosphorene to ~0.3 eV for bulk black phosphorus[52-55]. Up till now, the black phosphorus has been studied as a promising material for transistors[47,51,56,57], photodetectors[58], solar cells[59]. It is also considered as future material for thermoelectric devices[60,61], gas sensing[62] and electrodes in lithium ion batteries[63]. Moreover, the optical absorption of a few layer black phosphorus can be saturated by high intensity laser light. Up till now, the effect of saturable absorption in black phosphorus was shown at 532 nm and 1030 nm[64]. Taking into account the band-gap of black phosphorus and the fact that its absorption can be saturated it might serve as a broadband SA in the spectral range up to ~4 µm.

In this paper we experimentally demonstrate, that mechanically exfoliated ~300 nm-thick layers of black phosphorus deposited on the end-facet of a fiber connector might serve as a saturable absorber for ultrashort pulse generation in fiber lasers. We have characterized the nonlinear optical properties of the material as a function of input light polarization. The transmission of as-prepared SA was changed by a maximum value of 4.4% when the fluence of the radiation propagated in the fiber core reached 540 µJ/cm$^2$. Such effective modulation depth allows for mode-locked operation in fiber lasers based on Er-and Tm-doped fibers. The 272 fs-short pulses generated from the Er-doped fiber laser were centered at 1560.5 nm with 10.2 nm of full-width at half maximum (FWHM). The Tm-doped laser was capable of generating 739 fs pulses at 1910 nm with 5.8 nm FWHM. This is the first demonstration of ultrashort pulse generation using SA based on black phosphorus. The obtained results unambiguously demonstrate great potential of this new nanomaterial for future applications.

**Results and discussion**

The black phosphorus-based SA was prepared by mechanical exfoliation method extensively developed for graphene and other two-dimensional materials[15,23,24]. The obtained layers were transferred onto the fiber ferrule, so that they entirely cover the fiber core, what is clearly seen in the scanning electron microscope (SEM) image presented in the Figure.1a. Thanks to the visible boundary of the 125 µm fiber cladding, we could precisely locate the position of the core under the black phosphorous layer. The composition of the transferred layer was confirmed by the energy dispersive X-ray spectroscopy (EDX). The analysis of the spectroscopy data presented in Figure. 1b confirms that the transferred material is black phosphorus. The images taken from the fiber core area show only the presence of phosphorus, oxygen and silicone. The last two elements originate from the substrate which is the optical fiber made from high purity silicone dioxide (SiO$_2$). The morphology of the deposited layers was investigated by atomic force microscopy (AFM). The 50x50 µm$^2$ AFM image comprising the 81 µm$^2$ fiber core area (the diameter of the SMF-28 fiber core is 9 µm) is depicted in the Figure 1c. The highest thickness of the transferred layer in the investigated area was ~700 nm. However, the maximum thickness in the area which covers the core was around 300 nm. The cross section of the AFM image made through the fiber core shows that the sample was relatively uniform in this range.

The fiber connector with deposited black phosphorus layer was then connected with a clean one via a fiber adapter, forming a fiber-integrated SA device. Its transmittance in the 900 – 2200 nm spectral range measured using a low intensity broadband white light source is depicted in the Figure 2a. it changes from around 30% at 900 nm to 55% at 2200 nm. The oscillatory character of the measured transmittance correlates very well with the previously calculated optical conductivity for 20 nm thick black phosphorus and can be related to the underlying electronic subbands structure[65]. The power-dependent transmittance of the SA shown in the Figure 2b was measured using a high-intensity mode-locked laser operating at 1560 nm. Changing the incident fluence in the 10 – 550 µJ/cm$^2$ range (limited by the available pump source) the transmittance of black phosphorus was increased by maximum of 4.4%. Due to the fact, that black phosphorus is an anisotropic crystal, its linear absorption is sensitive to the incident light polarization state[55,65]. Hence, in order to investigate the impact of the polarization on the nonlinear parameters of black phosphorus, the polarization azimuth of the measurement beam incident on the SA was changed by a polarization controller. Similar to T.Low et al.[65] the linear absorption is polarization-dependent and can be maximally tuned by about ~11%. For all tested azimuths the black phosphorus layer exhibits saturable absorption effect (Figure 2b), but the modulation depth is polarization-dependent. The effective modulation depths (limited by the available excitation fluence) varies in the 0.6 – 4.4% range. The saturation fluence increases with losses, was also observed for graphene[28]. Comparing our results with that reported previously[65] the highest transmittances is obtained when the polarization is parallel to the y-axis of the black phosphorus lattice structure[55,65]. In that case full saturation of the SA was observed. Further increase of the input fluence leads to a roll-off in the transmittance curve, which is connected mainly with two-photon absorption (TPA) in the SA[66-68]. The measurement data are in very good agreement with the theoretical calculations done according to the saturable absorber model[67,68] (the olive lines in the Figure 2b). Based on the model and taking into account the measurement data where the roll-off effect was clearly observed (the two curves with higher transmittance) the TPA coefficient ($\beta_{TPA}$) of ~500 cm/MW was calculated. It is at the same level as bi-layer graphene $\beta_{TPA}$[69]. Taking into account the broadband absorption of black phosphorus and its nonlinear optical response leading to absorption saturation at relatively low fluence, it meets basic criteria of a passive SA for fiber lasers.

The as-prepared and characterized black phosphorus-based SA was used to generate ultrashort optical pulses in two ring cavities based on Er- and Tm-doped fibers. The general concept of the laser setup is presented in Figure 3. In order to obtain pulsed operation in mode-locked regime the lasers were pumped above the threshold and the intracavity polarization state was optimized by the polarization controller (PC). The mode-locked operation in the Er-doped fiber laser, pumped by a 980 nm laser diode, was obtained when the pump power reached the 80 mW threshold. Afterwards, the pumping power might be reduced to 45 – 65 mW in order to obtain stable, single-pulse operation without any continuous-wave (CW) component. The generated optical solitons containing the typical Kelly's sidebands with FWHM of 10.2 nm were centered at 1560.5 nm (Figure 4a). The pulses were slightly chirped as evidenced by the time-bandwidth-product (TBP), which was at the level of 0.34 for the obtained 272 fs pulse duration (Figure 4b). The measured radio frequency (RF) spectrum (Figure 4c) is free of any spectral modulations and indicates that the signal to noise ratio (S/N) was better than 65 dB. The repetition frequency, resulting from the length of the resonator was 28.2 MHz. The generated pulse train was stable without any signs of Q-switched mode-locking or multiple-pulse generation (Figure 4d). The average output power measured at 65 mW of pumping was at the level of 0.5 mW.

The Tm-doped fiber laser was pumped by a 1568 nm laser diode, beforehand amplified in an Erbium-Ytterbium co-doped fiber amplifier (EYDFA) to about 500 mW. The ultrashort pulses with the 5.8 nm FWHM centered at 1910.5 nm (Figure 5a) were obtained when the pumping power exceeded the threshold of 310 mW. The stable, single pulse operation was observed up to the pump power level of 395 mW. Pulse-breaking and parasitic CW components were observed at higher pump powers. Similar to the Er-doped laser, the generated 739 fs pulses were also slightly chirped, with the TBP at the level of 0.35 (Figure 5b). The S/N ratio of the generated pulses was better than 65 dB (Figure 5c). The RF spectrum and the corresponding oscilloscope trace (Figure 5d) indicate that the laser operation was free of parasitic spectral modulations. The pulse repetition rate was 36.8 MHz. The output power measured at 395 mW was 1.5 mW. The characteristic dips observed on the optical spectrum are related to the water absorption lines, which are densely located in the 1.9 μm region. In order to confirm it, we have simulated the absorption of light over 1 m path in air with 1% water content using HITRAN database[70]. The simulated absorption lines are plotted (with blue line) in the Figure 5a. It can be seen, that the location and amplitude of the absorption peaks on the spectrum ideally match the simulated water lines.

It is worth to emphasize the high optical damage threshold of the black phosphorus layer. During the performed experiments both lasers were pumped up to the 0.5 W and no signs of optical damage of the SA were observed. Moreover this first demonstration of ultrashort pulse generation using mechanically exfoliated black phosphorus presents significantly better results than the first reports on mode-locking with the use of carbon nanotubes[18] and graphene[12,13]. This demonstrates the grate potential of black phosphorus not only in electronics but also in photonics.

In conclusion, we have shown that the mechanically exfoliated layers of black phosphorus transferred onto a fiber might be used as a broadband saturable absorber for ultrashort pulse generation. We have for the first time experimentally demonstrated the saturable absorption effect of black phosphorus in the 1550 nm and 1900 nm spectral ranges. The prepared saturable absorber allows for ultrashort optical pulses generation in the fiber lasers based on Er- and Tm-doped active fibers. Generated pulses with 272 fs and 739 fs duration were centered at 1560.5 nm and 1910.5 nm, respectively. The obtained results clearly indicate that the number of applications of black phosphorus is still growing. Taking into account the unique electronic and optoelectronic properties, the global interest in this material can reach the same level as in graphene. Moreover, recently demonstrated fabrication process of large quantities of stable few-layer black phosphorus dispersions[64] can accelerate research on its new applications.

**Methods**

As a base material for SA preparation a commercially available high purity piece of black phosphorus (Smart Elements, purity 99.998%) was used. In the first step, black phosphorus flakes were obtained by precise scraping the lumps with a sharp blade. Then, the flakes were mechanically exfoliated using scotch tape. Afterwards, the thin layers were pressed with a fiber ferule (the FC/APC connector), previously ultrasonically cleaned in the presence of isopropyl alcohol and transferred onto the fiber core.

The fiber with deposited black phosphorus layer was investigated with AFM in tapping mode (Veeco diDimension V SPM with NanoScope V controller). The fiber was placed in a holder, in order to adjust the measured surface parallel to the microscope scan area. The measurement was done using Bruker MPP-11100-10 AFM cantilever with 0.1 Hz scan rate. The amplitude of the lever was set to 500 mV.

The position and quality of the black phosphorus layer transferred onto the fiber ferrule were visualized using scanning electron microscope (SEM, Hitachi SU 6600). To confirm unambiguously correct location of black phosphorus layer we utilize the SEM equipped with an energy dispersive X-ray Spectrometer (EDX) - NORAN System 7 X-ray Microanalysis System. Better visualization was obtain by using an advanced variable pressure technology in the SEM microscope. Despite the measured sample is non-conductive, it was not covered with metal coating. Nonetheless, the quality of obtained images is good enough to visualize the exfoliated layer placed on the fiber core. The composition of the exfoliated layer was confirmed to be pure phosphorus by EDS analysis.

Optical characterization of the black phosphorus-based SA was performed in the linear and nonlinear regime using a broadband white light source and a pulsed fiber laser, respectively. The linear transmittance of the SA was measured in the range from 900 nm to 2200 nm using white light source (Yokogawa AQ4305) and optical spectrum analyzers (Yokogawa AQ6370B, Yokogawa AQ6375). The measurement were performed with 1 nm spectral resolution and the highest available sensitivity. The nonlinear power-dependent transmission was investigated using an all-fiber setup (Figure 6), similar to that reported before[13,31] with femtosecond laser operating at 1560 nm used as an excitation source (Menlo Systems T-Light, 100 MHz, 100 mW). The power from the reference arm and passing through the measured sample was simultaneously measured by the optical power meter (Newport 2936-R with 918D-IG-OD3R heads). The average power incident on the sample was changed by the electrically controlled variable optical attenuator (VOA) with about 3 dB insertion losses. The laser beam was divided into reference and measurement arm by a 20/80 coupler. The pulse duration in the SA plane was 900 fs. The measurement data were fitted with a commonly used two-level model function of SA[67,68] including the TPA effect:

$$T(F) = \alpha_{ns} - \left(1 - exp\left(-\frac{F}{F_{sat}}\right)\right) \cdot \frac{\Delta\alpha \cdot F_{sat}}{F} - \frac{F}{F_2}$$

where $\alpha_{ns}$ – non-saturable losses: $\Delta\alpha$ – modulation depth, $F_{sat}$ – saturation fluence, $F_2$ - inverse slope of the induced absorption effect[68]. Taking into account the fitting functions where the roll-off effect is clearly observed (the last two with the highest transmittance) the $F_2$ parameter was estimated at the level of ~60 µJ/cm². According to the SA model[67] the $β_{TPA}$ coefficient was calculated as:

$$β_{TPA} ≈ \frac{τ_p}{F_2 \cdot d}$$

where: $τ_p$ – pulse duration, d - thickness of the absorbing layer (the exfoliated black phosphorus layer thickness was ~300 nm).

In order to investigate the influence of polarization azimuth on the power-dependent transmittance the polarization in the measurement setup depicted in Figure 6 was changed via PC. The polarization azimuth and polarization state in the plane of the SA was monitored using the polarimeter (Thorlabs PAX 5710) equipped with the measurement head (PAN5710IR3).

The general setup of the mode-locked fiber laser is depicted in Fig. 4. In both cases the same laser cavity configuration was used. As active fiber, a 40 cm long piece of LIEKKI™ Er80-8/125 (EDF) and 120 cm long piece of Nufern SM-TSF-9/125 (TDF) were used to generate laser pulses at 1550 nm and 1900 nm, respectively. The Er-doped fiber was counter-directionally pumped via a fused 980/1550 nm wavelength division multiplexer (WDM) by a 980 nm laser diode. In order to pump the Tm-doped fiber, an Er/Yb-doped fiber amplifier seeded by a 1568 nm laser diode was used. This laser setup was also counter-directionally pumped via a filter type 1570/2000 nm WDM. The signal was coupled out using 20% output couplers (OC) in both cases. Fiber-based in-line PC allows to adjust the intra-cavity polarization and start the mode-locked operation. The fiber isolator forced unidirectional signal propagation. Both investigated setups were designed in all-anomalous dispersion regime. The Er-doped laser resonator consists of 0.4 m EDF with group velocity dispersion (GVD, $β_2$) at the level of 59 ps²/km, 0.3 m HI1060 fiber with $β_2$= −7 ps²/km and 6.52 m SMF-28 fiber with $β_2$= −22 ps²/km, which results in the total net group delay dispersion (GDD) of approx. −0.122 ps². The Tm-doped laser resonator consists of 1.2 m TDF with $β_2$ at the level of −76 ps²/km and 4.35 m SMF-28 fiber with $β_2$= −68 ps²/km, which results in the total net GDD of approx. −0.387 ps². The performance of the lasers was observed using an optical spectrum analyzer (Yokogawa AQ6375), 12 GHz digital oscilloscope (Agilent Infiniium DSO91304A), 7 GHz RF spectrum analyzer (Agilent EXA N9010A) coupled with a 16 GHz photodetector (Discovery Semiconductors DSC2-50S), and optical autocorrelators (APE PulseCheck for 1550 nm and Femtochrome FR-103XL for 1910 nm).

**Acknowledgements**

The research was partially supported by: the statutory funds of the Chair of Electromagnetic Field Theory, Electronic Circuits, and Optoelectronics and the Division of Microelectronics and Nanotechnology, Wroclaw University of Technology under the grants no. S40043 and S40012, the National Centre for Research and Development through Applied Research Program grant no. 178782, the National Centre for Science (NCN, Poland) under grant no. DEC-2012/07/D/ST7/02583. The work on the Tm-doped fiber laser presented in this paper was supported by the NCN under the project "Passive mode-locking in dispersion-managed ultrafast Thulium-doped fiber lasers" (decision no. DEC-2013/11/D/ST7/03138).

**Authors contribution**

W.M. performed the SEM and EDX analysis of the exfoliated black phosphorus layers, P.P. performed the AFM measurements, K.M.A. discussed the obtained results, G.S. performed the nonlinear optical measurements, characterized the Tm-doped fiber laser performance and wrote the paper, J.S. conceived the study, planned the experiments, performed mechanical exfoliation process, measured the SA linear absorption, assembled the fiber lasers and wrote the paper.

**Figures**

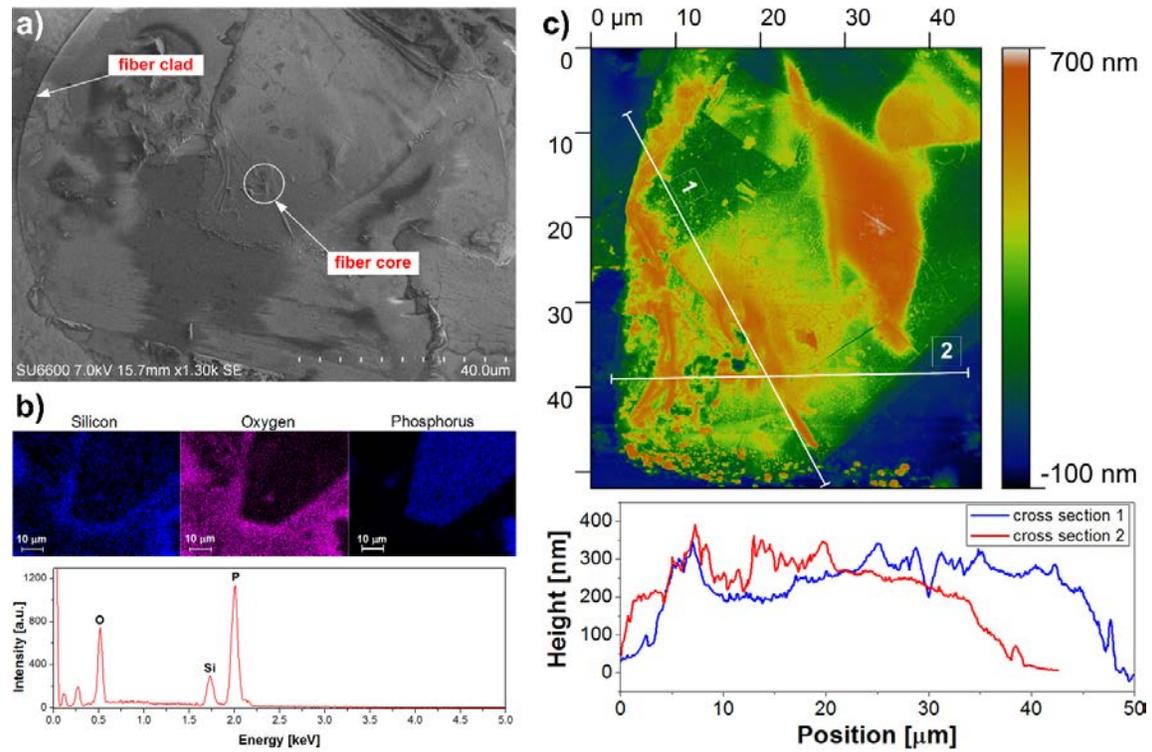

Fig. 1. Characterization of the exfoliated black phosphorus layers, transferred onto the fiber core: a) SEM image with marked fiber clad and core. It is clearly confirmed that exfoliated layer cover the fiber core, b) EDX spectroscopy data, c) AFM image including the cross section through the fiber core area

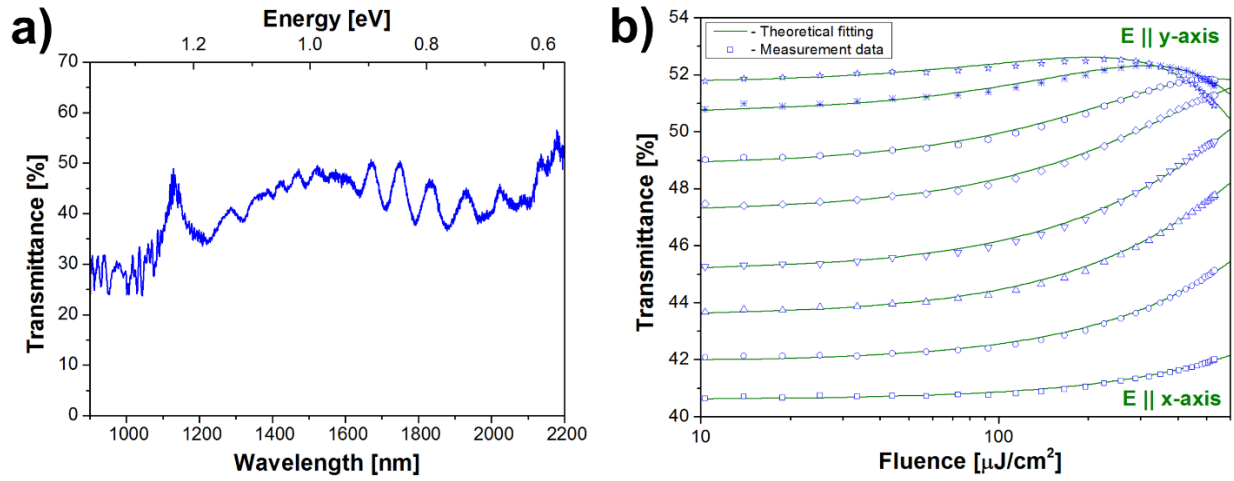

Fig. 2. The transmittance of the exfoliated black phosphorus layer measured using: a) low intensity (broadband white light), b) high intensity (mode-locked laser) light source. The power-dependent transmittance was measured for polarization azimuths changing form E ∥ x-axis to E ∥ y-axis

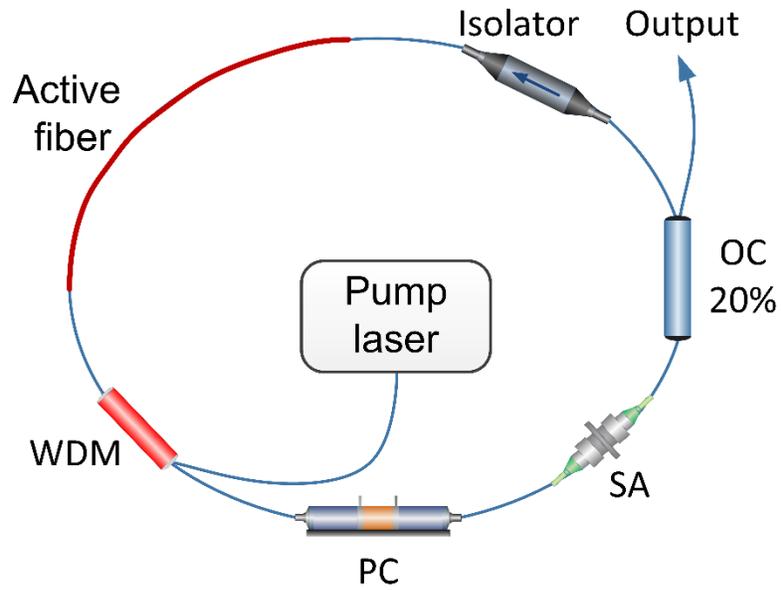

Fig. 3. Setup of a mode-locked fiber laser based on ring cavity architecture (WDM – wavelength division multiplexer, OC – output coupler, SA – saturable absorber, PC – polarization controller)

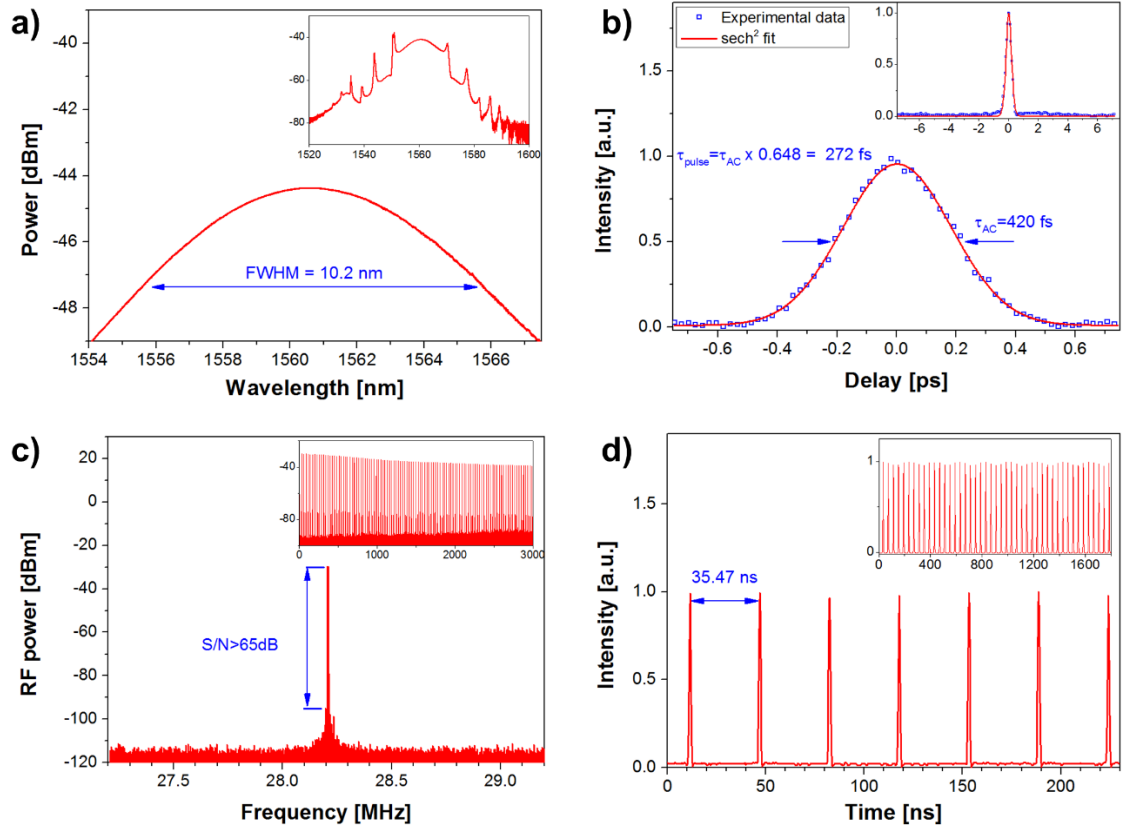

Fig. 4. Performance of the Er-doped mode-locked fiber laser measured at the pump power of 65 mW: a) optical spectrum with the FWHM of 10.2 nm (measured with 0.05 nm resolution), b) autocorrelation trace of the 272 fs pulse, c) RF spectrum measured in the 2 MHz span with 60 Hz RBW presenting the repetition frequency of the laser and the S/N ratio. Inset: the broad range of harmonics without spectral modulations measured in the 3 GHz span with 47 kHz RBW, d) oscilloscope trace

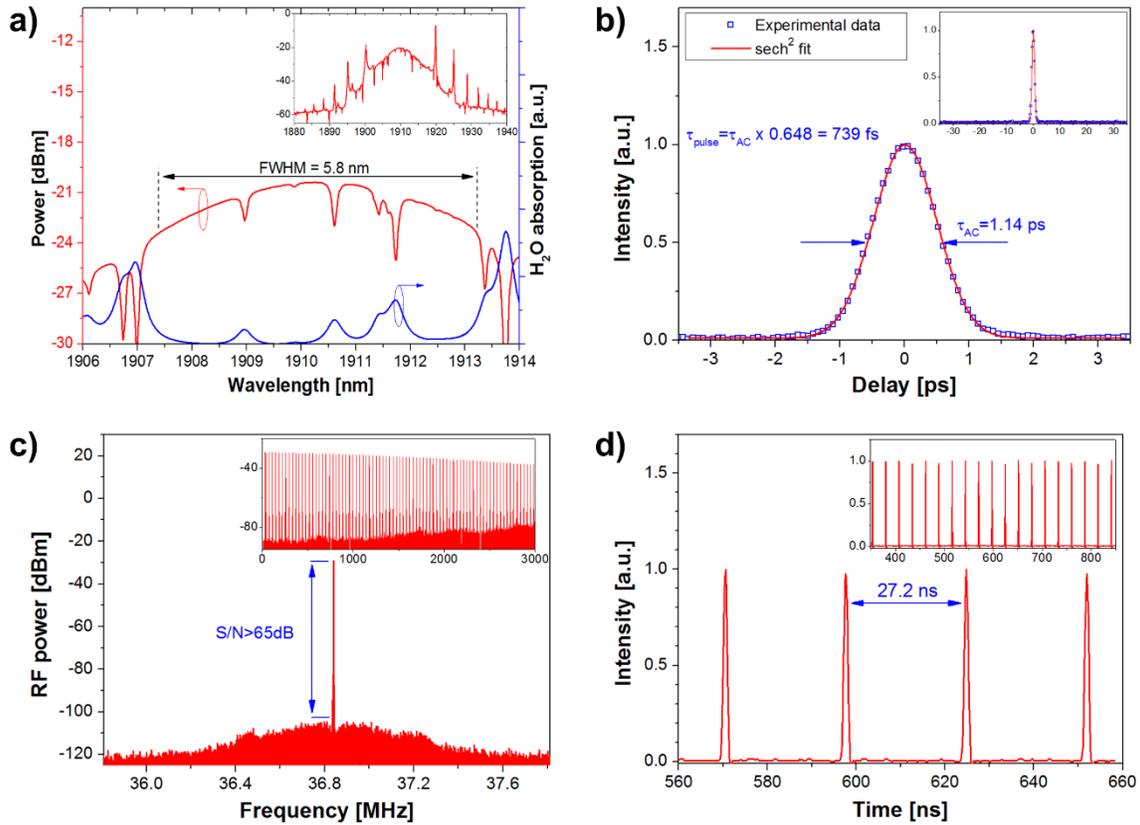

Fig. 5. Performance of the Tm-doped mode-locked fiber laser measured at the pump power of 395 mW: a) optical spectrum with the FWHM of 5.8 nm measured with 0.05 nm resolution (the blue line represents the water absorption lines in air), b) autocorrelation trace of the 739 fs pulse, c) RF spectrum measured in the 2 MHz span with 80 Hz RBW presenting the repetition frequency of the laser and the S/N ratio. Inset: the broad range of harmonics without spectral modulations measured in the 3 GHz span with 200 kHz RBW, d) oscilloscope trace

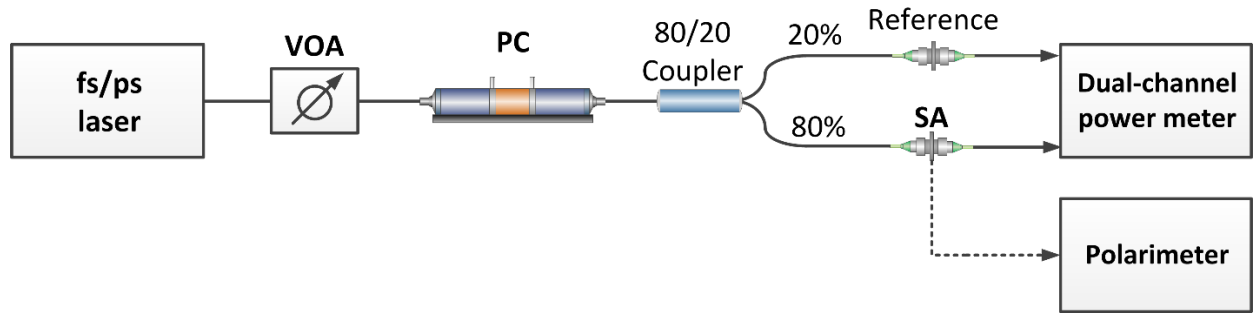

Fig. 6. The setup of the power-dependent transmittance measurement. The polarization azimuth and polarization state was controlled and monitored by the PC and polarimeter, respectively.